\documentclass[preprint,nofootinbib,nobibnotes,groupaddress,preprintnumbers,aps,prd]{revtex4}
\usepackage{bm,epsfig,mathrsfs,amsmath,amssymb,graphicx}
\usepackage{epsfig,amsmath,graphicx,amssymb}
\usepackage{graphicx}
\usepackage{dcolumn}
\usepackage{bm}
\def\dbar{{\mathchar'26\mkern-12mu d}}

\begin{document}

\title{Efficiency at maximum power output of an irreversible Carnot-like cycle with internally dissipative friction}

\author{Jianhui Wang $^{1, 2}$}  \email{physwjh@gmail.com}
\author{Jizhou He $^1$}
 \affiliation{ $^1\,$ Department of
Physics, Nanchang University, Nanchang 330031, China \\
$^2\,$ State
Key Laboratory of Surface Physics and Department of Physics, Fudan
University, Shanghai 200433, China}

\begin{abstract}

We investigate the efficiency at maximum power of an irreversible
Carnot engine performing finite-time cycles between two reservoirs
at temperatures $T_h$ and $T_c$ $(T_c<T_h)$, taking into account of
internally dissipative friction in two ``adiabatic" processes. In
the frictionless case, the efficiencies at maximum power output are
retrieved to be situated between $\eta_{_C}/$ and
$\eta_{_C}/(2-\eta_{_C})$, with $\eta_{_C}=1-T_c/{T_h}$ being the
Carnot efficiency. The strong limits of the dissipations in the hot
and cold isothermal processes lead to the result that the efficiency
at maximum power output approaches the values of $\eta_{_C}/$ and
$\eta_{_C}/(2-\eta_{_C})$, respectively. When dissipations of two
isothermal and two adiabatic processes are symmetric, respectively,
the efficiency at maximum power output is founded to be bounded
between $0$ and the Curzon-Ahlborn (CA) efficiency
$1-\sqrt{1-\eta{_C}}$, and the the CA efficiency is achieved in the
absence of internally dissipative friction.


PACS number(s): 05.70.Ln, 05.30.-d
\end{abstract}

\maketitle
\date{\today}

\section {introduction}
Quasistatic Carnot cycle is the most efficient heat engine cycle
allowed by physical laws. Practically any heat engine operates far
from the ideal maximum efficiency conditions set by Carnot
\cite{Car24}. Although the  Carnot cycle has the highest efficiency,
its power output is zero because the time for completing a cycle is
infinite. The cycle should be speeded up to obtain a finite power.
Considering a finite-time Carnot cycle within the assumption of
endoreversibility that irreversible processes occur only through the
heat exchanges, Curzon and Ahlborn (CA) \cite{Cur75} obtained the
efficiency  at maximum power output as
\begin{equation}
\eta_{_{CA}}=1-\sqrt{{T_c}/{T_h}}, \label{etth}
\end{equation}
where $T_h$ and $T_c$ are the temperatures of the hot and cold heat
reservoirs, respectively.  Historically  speaking, the seminal
expression (\ref{etth}) was derived by Yvon \cite{Yvo55} and Novikov
\cite{Nov58} much earlier than \cite{Cur75}. But it is usually
called the CA efficiency. Recently,  the issue of the efficiency at
maximum power output, as main focus of finite time thermodynamics,
has attracted much interest \cite{ Che89, Esp10, Esp09, Esp11,
Ape12, Oue12, pre11, pre12I, pre12II, Gor91, Sch08, Tu08, Tu12,
Gav10, Mor12, Van05, Izu08,  Gev92, Fel00, Rez06, Abe11, Wang11,
Chen11, Sei11, Tu48}. Under the low-dissipation assumption that the
irreversible entropy production in a heat-exchange process is
inversely proportional to the time required for completing that
process, Esposito \emph{et al}. \cite{Esp10} proposed a model for
low-dissipation Carnot-like engines, in which use of
endoreversibility hypothesis and phenomenological transfer laws can
be avoided .

Although the importance of internally frictional dissipation in an
adiabatic process was mentioned by Novikov in his pioneer paper
\cite{Nov58,note1}, most of the studies about the efficiency at
maximum power output always neglect the influence of inner friction
on the performance of the heat engine models, within the assumption
that the time taken for completing the adiabatic process is ignored
or proportional to the total time spent on the isothermal processes.
  From
everyday experience, the irreversible phenomena that limits the
optimal performance of engines  occurs not only in an isothermal
process but but also in an adiabatic process because of inner
friction when classical or quantum piston moves \cite{Jar12, Nak11,
Fel00, Gor91, Rez06, Biz12}. Dissipation loss due to internally
dissipative friction, by which real engines are dominated
\cite{Gor91},
 has been discussed in several papers \cite{Fel00, Wang07, Rez06, Wu92,
pre12I}. However, so far there has been no comprehensive discussion
of the effects of friction on the cycle performance in the
literature, and thus the properties of an irreversible Carnot-like
cycle consisting of two irreversible isothermal and two
non-isentropic adiabatic processes have not been addressed
adequately and clearly. For this reason, we follow the tradition of
thermodynamics constructing a more generalized engine, in which the
``adiabatic" process  takes finite time as well as becoming
non-isentropic \cite{Wang07, Fel00, Rez06, Nak11, adia}. Troughout
this paper, the word ``isothermal"  merely indicates that the
working substance is in contact with a reservoir at constant
temperature, and the word ``adiabatic"  means that the working
substance is isolated from a heat reservoir and no heat exchange
happens.

In this paper,  we focus on the study of the efficiency at maximum
power output of an irreversible Carnot-like  engine  performing
finite time cycles, in which frictional dissipation and the time of
any adiabat are taken into account. We derive the cycle period which
contains time spent on four thermodynamic processes, which is quite
different from that derived in the previous models for which the
time of two adiabatic processes was assumed to either be
proportional to the time duration of the two isothermal processes or
be negligible.  In the frictionless case, the efficiencies at
maximum power output are  proved to bounded between $\eta_{_C}/$ and
$\eta_{_C}/(2-\eta_{_C})$, with  the Carnot efficiency
$\eta_{_C}=1-T_c/{T_h}$, coinciding with the result found in Ref.
\cite{Esp10} for frictionless engine models in which the time
required for completing any adiabatic process was assumed to be
totally ignored. In the strong limits of the dissipations in the hot
and cold isothermal processes, we find that the efficiency at
maximum power output approaches the values of $\eta_{_C}/$ and
$\eta_{_C}/(2-\eta_{_C})$, respectively. If dissipations of two
isothermal and two adiabatic processes are symmetric, respectively,
we prove that the efficiency at maximum power output is bounded
between $0$ and the Curzon-Ahlborn (CA) efficiency
$1-\sqrt{1-\eta{_C}}$, and that the CA efficiency is  reached by
ignoring the friction.

\section {Engine model}

\begin{figure}[h]
\includegraphics[width=250pt]{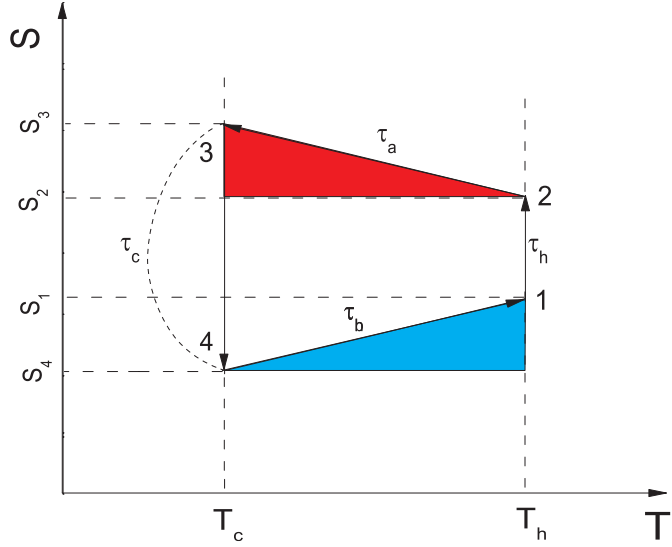}
\caption{ (Color online) Schematic diagram of an irreversible
Carnot-like cycle in the plane of the temperature $T$ and entropy
$S$. The values of the entropy $S$ at the four special instants are
indicated. $t_h$ ($t_c$) is the time allocation when in contact with
the hot (cold) reservoir. $t_a$ represents the time allocation for
adiabatic   expansion and $t_b$ for adiabatic compression.  The red
triangular area is the heat developed by friction for the process
$2\rightarrow3$, while the blue triangular area represents the
additional heat for the process $4\rightarrow1$.} \label{st}
\end{figure}

A Carnot-like cycle
$1\rightarrow2\rightarrow3\rightarrow4\rightarrow1$ is drawn in the
$(S, T)$ plane (see Fig. \ref{st} ).  During two isothermal
processes $1\rightarrow2$ and $3\rightarrow4$, the working substance
is coupled to a hot and a cold heat reservoir at constant
temperatures $T_h$ and $T_c$, respectively.  Let $S_i$ be the
entropies at the instants $i$ with $i=1,2,3, 4$.  For the reversible
cycle where $S_2=S_3$ and $S_1=S_4$, we recover the Carnot
efficiency $\eta_{_C}=1-\frac{T_c}{T_h}$, which is independent of
the properties of the working substance. In the adiabatic process
$2\rightarrow3$ ($4\rightarrow1$), the working substance is
decoupled from the hot (cold) reservoir, and the entropy changes
from $S_2$ to $S_3$ ($S_4$ to $S_1$) in a period $t_{a}$ ( $t_{b}$).

Let us consider the Carnot-like cycle under finite-time operation.
 Finite-time  cycles  move
the working substance away from the equilibrium, leading to
irreversibility of the engine. Although the system needs no close to
equilibrium during the isothermal process,  the system remains in an
equilibrium state with the heat reservoir at the special instants
$i$ where $i=1, 2, 3, 4$. Under such a circumstance,  the
thermodynamic quantities of the system$-$in particular the
entropy$-$are well defined at these instants. Unlike in the
frictionless case where any adiabatic process is isentropic, the
adiabatic process becomes non-isentropic when friction is included,
since friction develops heat and leads to an increase in entropy in
any adiabat. This additional heat remains in chamber or in  trap
along an adiabatic process until it is released to a heat reservoir
with which the working systems couples during an isothermal process.
As a consequence, heat productions due to friction in the adiabatic
expansion $2\rightarrow3$ and  in the adiabatic compression
$4\rightarrow1$ are released into the cold and hot reservoirs,
respectively \cite{Ape12, Wu92}. This additional  heat is also
represented in Fig. \ref{st} by the red triangular area for the
branch $2\rightarrow 3$ and by the blue triangular area for the
process $4\rightarrow1$. Note that, the heat produced during the
process $4\rightarrow 1$ decreases the absorbed heat during the hot
isothermal process, and the heat produced during the adiabat
$2\rightarrow 3$, as pure loss, is released to the cold reservoir.
The cycle model is operated in the following processes.

\emph{1. Isothermal expansion $1\rightarrow2$} . At time $t=0$, the
working substance is brought into contact with a hot reservoir at
constant temperature $T_h$. The hot reservoir is then removed after
time duration $t_{h}$.  An amount of heat absorbed from the
surroundings is represented by $Q_h$. In this process, the entropy
is changing from the initial entropy $S_1$ to the entropy $S_2$. The
entropy variation, $\Delta S_h=S_2-S_1$, is given by
\begin{equation}
\Delta S_h=Q_h/{T_h}+\Delta S^{ir}_h,
\end{equation}
where $\Delta S^{ir}_{h}$ is the irreversible entropy production.

\emph{2. Adiabatic expansion $2\rightarrow3$}. The working substance
decouples from the hot reservoir for a time duration $t_{a}$. In
this process the frictional dissipation develops heat, resulting in
the fact that the entropy increases from $S_2$ to $S_3$. The entropy
production arising from the inner friction is denoted by
\begin{equation}
\Delta S_a= S_3-S_2=\Delta S^{ir}_a.
\end{equation}

\emph{3.  Isothermal compression $3\rightarrow4$}. The  working
substance is coupled to a cold reservoir at constant temperature
$T_c$ for time $t_c$. Then entropy changes on this process from
$S_3$ to the entropy $S_4$. For the cycle to close, $S_4$ should be
smaller than $S_1$.  The variation of the entropy can be expressed
as
\begin{equation}
\Delta S_c =|S_3-S_4|=Q_c/{T_c}+\Delta S^{ir}_c,
\end{equation}
where  $Q_c$ is the amount of heat released directly to the cold
reservoir, and $\Delta S^{ir}_{c}$ is the irreversible entropy
production.

\emph{4. Adiabatic compression $4\rightarrow1$}.  Similar to the
adiabatic expansion, the working substance decouples from cold
reservoir. The time required for completing this process reads
$t_b$. The entropy increases from $S_4$ to the original value $S_1$.
The amount of entropy production due to the internally frictional
dissipation during this process is given by
\begin{equation}
\Delta S_b=S_1-S_4=\Delta S_b^{ir}. \label{deir}
\end{equation}

 After performing a whole cycle, the
system recovers the initial state and thus its  total energy remains
unchanged for the whole cycle. The work output after a single cycle
can be expressed as
\begin{equation}
W=Q_h-Q_c={(T_h-T_c)}\Delta S-T_h(\Delta
S^{ir}_h+\frac{S^{ir}_b}2)-T_c (\Delta S^{ir}_c+\frac{S^{ir}_a}2),
\label{wqrb}
\end{equation}
where  $\Delta S\equiv S_2-S_1$ is a state variable only depending
on the initial and final states of the isothermal processes,
whereas, $\Delta S^{ir}_h$, $\Delta S^{ir}_c$, $\Delta S^{ir}_a$,
and $\Delta S^{ir}_b$ are process variables depending on the
detailed protocols. 

\section{Efficiency at maximum power output}

 To continue our analysis, we denote, by min$\left\{\Delta
S^{ir}_\kappa\right\}\equiv L_\kappa(t_\kappa)$ \cite{Sch08, Esp10,
Gor91, Fel00, Tu12, Bli12, Wang07, Tuar} with $\kappa=h,c, a,$ and
$b$, the minimum irreversible entropy production  for the optimized
protocols. Physically,  the larger time required for completing the
corresponding process, the closer the process is to quasistatic
process, indicating that the irreversible entropy productions
$S_\kappa$ become much smaller and tend to be zero in the longtime
limits ($t_\kappa\rightarrow \infty)$. In other words,
$L_\kappa(t_\kappa)$ should be a monotonous decreasing function of
$t_\kappa$ for a given process $\kappa$. Because the irreversible
entropy production $L_\kappa$ is a function of the time $t_\kappa$
for a given process $\kappa$,   the irreversible entropy production
$ L_a(t_a)$ [$L_b(t_b)$] in the adiabatic process $2\rightarrow 3$
($4\rightarrow 1$) can not be included by the irreversible entropy
production $L_c(t_c)$ [$L_h(t_h)$] during the cold (hot) isothermal
process.

For convenience, we  make a variable transformation $x_\kappa =
1/{t_\kappa}$ with $\kappa=h, c, a$ and $b$, and  have the cycle
time $\tau=t_h+t_c+t_a+t_b=1/x_h+1/x_c+1/x_a+1/x_b$. Accordingly,
the power output $P=W/\tau$ and the efficiency $\eta=W/Q_h$ are
\begin{equation}
P=\frac{(T_h-T_c)\Delta S-T_h [L_h(x_h)+L_b(x_b)/2]-T_c \left[L_c
(x_c)+L_a (x_a)/2\right]}{(1/x_h+1/x_c+1/x_a+1/x_b)}, \label{pf2r}
\end{equation}
and
\begin{equation}
\eta=\frac{{(T_h-T_c)}\Delta S-T_h \left[L_h (x_h)+L_b
(x_b)/2\right]-T_c \left[L_c (x_c)+L_a
(x_a)/2\right]}{T_h\left[\Delta S- L_h (x_h)-L_b(x_b)/2\right]},
\label{et2t}
\end{equation}
respectively.

To specify the time allocation at maximum power output, the values
of $x_\kappa$, with $\kappa=h,c, a$ and $b$,  should be optimized.
We optimize power output $P$ over the time variables $x_\kappa$ to
obtain the time allocation during a cycle and thus to determine the
corresponding efficiency. Setting the derivatives of $P$ with
respect to $x_\kappa$ ($\kappa=h,c,a, b$) equal to zero, we derive
the following equations:
\begin{equation}
T_h L'_h x_h^2=W,  \label{th2w}
\end{equation}
\begin{equation}
T_c L'_c x_c^2=W, \label{tc2w}
\end{equation}
\begin{equation}
T_c L'_a x_a^2=2W, \label{ta2w}
\end{equation}
\begin{equation}
T_h L'_b x_b^2=2W, \label{tb2w}
\end{equation}
where the work $W$ was defined in Eq. (\ref{wqrb}), and
$L'_\kappa=\frac{d L_\kappa}{d x_\kappa}$ with $\kappa=h, c, a, b$.
Dividing Eq. (\ref{th2w}) by Eqs. (\ref{tc2w}), (\ref{ta2w}), and
(\ref{tb2w}), respectively, we find the optimal time allocation at
maximum power output:
\begin{equation}
\frac{x_c}{x_h}=\sqrt{\frac{T_h L'_h}{T_c L'_c}},  \label{xcxh}
\end{equation}
\begin{equation}
\frac{x_a}{x_h}=\sqrt{\frac{2T_h L'_h}{T_c L'_a}}, \label{xaxh}
\end{equation}
\begin{equation}
\frac{x_b}{x_h}=\sqrt{\frac{2 L'_h}{L'_b}}. \label{xbxh}
\end{equation}

Now we turn to the low-dissipation case where one assumes
$L'_\kappa(x_\kappa)=\Sigma_\kappa $ with $\Sigma_\kappa$ being
dissipation constants. In this case, we find, by substituting  Eqs.
(\ref{xcxh}), (\ref{xaxh}), and (\ref{xbxh}) into the maximization
condition $\frac{\partial P}{\partial x_h}=0$,  the physical
solution at
\begin{equation}
x_h=\frac{\Delta S(T_h-T_c)}{2 T_h\Sigma_h
\left(1+\sqrt{\frac{T_c\Sigma_c}{T_h\Sigma_h}}\right)}, \label{xhht}
\end{equation}
\begin{equation}
x_c=\frac{\Delta S(T_h-T_c)}{2 T_c\Sigma_c
\left(1+\sqrt{\frac{T_h\Sigma_h}{T_c\Sigma_c}}\right)}, \label{xcht}
\end{equation}
\begin{equation}
x_a=\frac{\Delta S(T_h-T_c)}{ T_c\Sigma_a
\left(\sqrt{\frac{2\Sigma_c}{\Sigma_a}}+\sqrt{\frac{2T_h\Sigma_h}{T_c\Sigma_a}}\right)},
\label{xaht}
\end{equation}
\begin{equation}
x_b=\frac{\Delta S(T_h-T_c)}{ T_h\Sigma_b
\left(\sqrt{\frac{2\Sigma_h}{\Sigma_b}}+\sqrt{\frac{2T_c\Sigma_c}{T_h\Sigma_b}}\right)}.
\label{xbht}
\end{equation}
The expressions (\ref{xhht}) and (\ref{xcht}) of the times spent on
a hot and cold isothermal process for a frictional heat engine are,
respectively, identical to corresponding ones [Eq. (7)] obtained a
engine model \cite{Esp10} in which both the internally dissipative
friction and the time taken for two adiabats were assumed to be
zero.

 Using Eq. (\ref{et2t}), together with  Eqs. (\ref{xhht}),
(\ref{xcht}), (\ref{xaht}), and (\ref{xbht}), it follows that
 the efficiency at maximum power becomes
 \begin{equation}
 \eta^*=\frac{\eta_{_C} \left(1+\sqrt{\frac{T_c\Sigma_c}{T_h\Sigma_h}}-\sqrt{\frac{T_c\Sigma_a}{2T_h\Sigma_h}}
 -\sqrt{\frac{\Sigma_b}{2\Sigma_h}}\right)}{1+2\sqrt{\frac{T_c\Sigma_c}{T_h\Sigma_h}}+\frac{T_c}{T_h}-
 \sqrt{\frac{\Sigma_b}{2\Sigma_h}}\left(1-\frac{T_c}{T_h}\right)}, \label{etht}
 \end{equation}
which is one of our main results in the paper.  Unlike the
frictionless cycle where the positive work condition is $T_c<T_h$,
the frictional cycle produces  positive work under the conditions
that $T_c<T_h$ and
\begin{equation}
\sqrt{\Sigma_a}+\sqrt{\frac{T_c}{T_h}}\sqrt{\Sigma_b}<\sqrt{2}(\sqrt{\Sigma_h}+\sqrt{\frac{T_c}{T_h}}\sqrt{\Sigma_c}).
\label{sqac}
\end{equation}
 Only when this positive work
condition is satisfied can the positive work be extracted.  We
present the efficiency at maximum power output in a broader context
by taking into account of inner friction and the time taken for any
adiabat.

 (1) When the inner friction is neglected,
i.e., $\Sigma_a\rightarrow0$ and $\Sigma_b\rightarrow0$, the
expression (\ref{etht}) of efficiency at maximum power output is
reduced to that from either stochastic thermodynamics \cite{Sch08}
 or by low-dissipation assumption \cite{Esp10}.  In  this frictionless case, the
completely asymmetric limits  $\frac{\Sigma_c} {\Sigma_h}\rightarrow
0$ and $\frac{\Sigma_c} {\Sigma_h}\rightarrow \infty$, causing the
efficiency $\eta_m$ at
 maximum power to approach the upper and lower bounds at
$\eta_+\equiv\frac{\eta_{_C}}{2-\eta_{_C}}$ and
$\eta_-\equiv\frac{\eta_{_C}}2$, respectively. That is, when
$\Sigma_a\rightarrow0$ and $\Sigma_b\rightarrow0$, the result of the
the bounds of the efficiency at maximum power for frictionless
Carnot-like cycle is retrieved,
\begin{equation}
\frac{\eta_{_C}}2\equiv\eta_-\leq\eta_m\leq\eta_+\equiv\frac{\eta_{_C}}{2-\eta_{_C}}.
\label{bounds}
\end{equation}
For   the symmetric dissipation ${\Sigma_c} ={\Sigma_h},$  the time
allocation to the hot and cold processes  satisfies
\begin{equation}
\frac{t_{h}}{t_{c}}=\sqrt{\frac{T_h}{T_c}}, \label {frtc}
\end{equation}
from which we can arrive at the CA efficiency
$\eta^*=\eta_{_{CA}}=1-\sqrt{\frac{T_c}{T_h}}$  by using Eq.
(\ref{etht}). We would like to emphasize that the results  are
identical to those in Ref. \cite{Esp10},  but they are derived in
the generalized engine model in which the  times required for
completing two adiabatic processes in the Carnot-like cycle are
obtained as Eqs. (\ref{xaxh}) and (\ref{xbxh}).

(2) The values of  frictional dissipations both $\Sigma_a$ and
$\Sigma_b$ are nonzero but finite. In  the  limits $
{\Sigma_h}\rightarrow \infty$ and ${\Sigma_c} \rightarrow \infty$,
the efficiency at maximum power output $\eta^*$ in Eq. (\ref{etht})
converges to the upper bound $\eta_+=\frac{\eta_{_C}}{2-\eta_{_C}}$
and to the lower bound $\eta_-=\frac{\eta_{_C}}2$, respectively.
Here the lower and upper bounds  are identical to the corresponding
ones in previous studies, but extended to the irreversible
Carnot-like cycle in which any adiabatic process is irreversible
because of internally frictional dissipation and its time is not
negligible.
\begin{figure}[h]
\includegraphics[width=300pt]{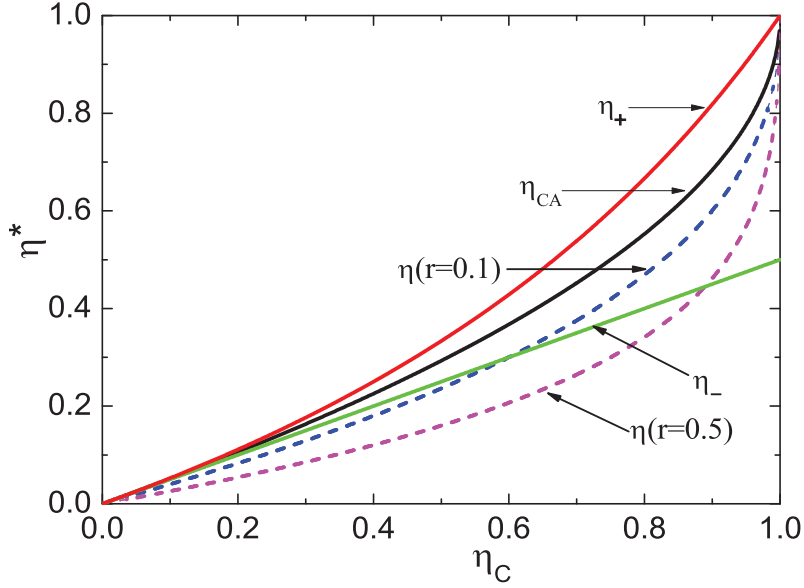}
\caption{ (Color online) Efficiency $\eta^*$ at maximum power as a
function of the Carnot value $\eta_{_C}$. The upper  and lower
bounds of efficiency at maximum power output in the frictionless
case $\eta_{+}$ and $\eta_{-}$, the CA efficiency $\eta_{_{CA}}$,
and the efficiencies for $r=0.1$ and $r=0.5$, are indicated in the
figure.} \label{etam}
\end{figure}

(3) Dissipations in two isothermal and two adiabatic processes are
symmetric, respectively, $\Sigma_h=\Sigma_c=r\Sigma_a=r\Sigma_b=r
\Sigma$ with $r$ and $\Sigma$ being constants. Then the efficiency
at maximum power becomes
\begin{equation}
\eta^*=\frac{\eta_{_C}{(\sqrt{r/2}-1)\left (1+ \sqrt{1 -
\eta_{_C}}\right)}}{-2\left (1+ \sqrt{1 -
\eta_{_C}}\right)+\eta_{_C}(\sqrt{r/2}+1)}, \label{et21}
\end{equation}
which satisfies the relation
\begin{equation}
0<\eta^*\leq  \eta_{_{CA}}=1-\sqrt{1-\eta_{_C}}. \label{0eac}
\end{equation}
Compared with the bounds (\ref{bounds}) for frictionless case, both
the upper and lower bounds (\ref{0eac}) of the efficiency at maximum
power are lowered when friction is introduced. Physically, this
originates from the fact that dissipative work is done to overcome
the inner friction which generates heat.  When
$r\rightarrow\sqrt{2}$, no positive work is extracted from the
cycle,thereby indicating that the efficiency becomes equal to zero.
From Eqs. (\ref{xcxh}), (\ref{xaxh}), and (\ref{xbxh}), we find that
the times spent on the four quantum thermodynamic processes are
distributed in such a way that
\begin{equation}
{t_h}/{t_c}={t_b}/{t_a}=\sqrt{{T_h}/{T_c}},
\end{equation}
with ${t_h}/{t_b}=\sqrt{r}$.  It should be noted that the efficiency
at maximum power increases as $r$ decreases, approaching the CA
efficiency $\eta_{_{CA}}$ in the frictionless limit ($r\rightarrow
0$).  In Fig. \ref{etam} we plot the efficiency (\ref{et21}) as a
function of $\eta_{_C}$ for $r=0.1$ and $r=0.5$, comparing
$\eta_{_{CA}}$  with the upper and lower bounds $\eta_{+}$ and
$\eta_{-}$ of frictionless engine models.

\section{Conclusions}

In conclusion, we have determined the efficiency at maximum power
for an irreversible Carnot-like engine which performs finite-time
cycles with internally dissipative friction.  In the limits of
extremely asymmetric dissipation
 ($\frac{\Sigma_c}{\Sigma_h}\rightarrow 0$ and
$\frac{\Sigma_c}{\Sigma_h}\rightarrow \infty$, with
$\frac{\Sigma_a}{\Sigma_h}\rightarrow 0$ and
$\frac{\Sigma_b}{\Sigma_h}\rightarrow 0$), the efficiency at maximum
power output converges to an upper and a lower bound
$\frac{\eta_{_C}}{2-\eta_{_C}}$ and $\frac{\eta_{_C}}2$, coinciding
with the result obtained previously in  the frictionless engine
model in which the time taken for two adiabatic processes was
ignored. When the
 dissipation in the hot (cold) isothermal process approaches the strong limit, i.e.,
 $\Sigma_h\rightarrow \infty$ ($\Sigma_c\rightarrow\infty$),
 the efficiency at maximum power output tends to be the upper
 (lower)
bound $(\frac{\eta_{_C}}{2-\eta_{_C}})$ ($\frac{\eta_{_C}}2$). When
the dissipations in two isothermal and two adiabatic processes are
symmetric, respectively,  we find that the efficiency at maximum
power output is bounded from above by the CA efficiency
$\eta_{_{CA}}$ and from below by zero, and that $\eta_{_{CA}}$ is
reached in the frictionless limit.

\emph{Acknowledgements:} We gratefully acknowledge support for this
work by the National Natural Science Foundation of China under
Grants No. 11147200 and No. 11065008, and the Foundation of Jiangxi
Educational Committee under Grant No. GJJ12136. J. H. Wang is also
grateful for  helpful discussions with Yann Apertet and Zhanchun Tu.


\begin{thebibliography}{99}
\bibitem{note1} In his paper \cite{Nov58}, Novikov pointed out the
importance of the entropy produced by internal irreversibilities
during two adiabatic processes, though he considered only an
endoreversible model by assuming that the two adibatic processes
were isentropic.
\bibitem{adia}  For a quantum system, the
first thermodynamic law is given by $d E=\dbar Q+\dbar W$, where
$\dbar Q=\sum_{n}E_ndp_n$, and $\dbar W=\sum_{n}p_ndE_n$. Here $P_n$
is the mean occupation probability and $E_n$ is the eigenenergy of
the $n$th eigenstate. According to  the quantum adiabatic theorem
\cite{Foc28}, an isolated system would remain in its initial state
during an aidabat. In order for the adiabatic theorem to apply, the
time scale of the change of the system state must be much larger
than that of the dynamical one, $\sim\hbar/E$ \cite{Foc28, pre12I,
pre12II, Abe11, pre11}. It is thus  indicated that the time required
for completing a quantum adiabatic process should be very large.
Otherwise, nonadiabatic dissipation (e.g., inner friction
\cite{Fel00, Wang07, Rez06, pre12I}) occurs because of rapid change
in the energy level structure of the quantum system and the quantum
adibatic condition is not satisfied. On the other hand, for a
classical system, the occupation probabilities can be varied during
a
 classical adiabatic process. For example,
when the process proceeds  fast,  but there is no heat exchange
between the working substance and the  heat bath.  The process is
classical adiabatic but not quantum adiabatic.  That is,  a quantum
adiabatic process is  a subset of a classical adibatic process, but
the inverse is not valid.
\end{thebibliography}
\end{document}